\documentstyle[12pt,epsf]{article}
\begin{document}
\setcounter{page}{1}
\begin{flushright}
{\bf LPTHE-ORSAY Preprint 94-84}\\
September 1994
\end{flushright}
\newcommand{\np}{Nucl. Phys. {\bf B}\ }
\newcommand{\pl}{Phys. Lett. {\bf B}\ }
\newcommand{\prd}{Phys. Rev. {\bf D}\ }
\newcommand{\zp}{Zeit. Phys. {\bf C}\ }
\newcommand{\prl}{Phys. Rev. Lett.\ }
\vskip 15 mm
\noindent
{\Large\bf Fermion masses and mixing from an extra gauge symmetry.}
\vskip 10 mm
\noindent
{\bf E. Papageorgiu}
\footnote{Talk given at the G\"ursey Memorial Conference on
Strings and Symmetries I, June 6-10, 1994, Istanbul, Turkey.}
\vskip 2 mm
\noindent
{Laboratoire de Physique Th\'{e}orique et Hautes Energies}
\noindent
{Universit\'{e} de Paris XI, B\^{a}timent 211, 91405 Orsay, France.}
\vskip 15 mm

\noindent
Looking for simplicity could mean looking for a symmetry.
For Feza G\"ursey this was more than a hypothesis or
just an attitude. It was a rewardfull commitment he shared with other
physicists.
Following the same attitude one may hope to find
an underlying symmetry principle for the long list of mass parameters
which characterize the elementary particle spectrum and
enter in the Standard Model (SM) as free parameters. To start with,
one can look for regularities in terms of some observable.
We know for example that the observed mass and mixing
hierarchies of the fermion spectrum can be described in terms of the
{\it Wolfenstein} parameter $\lambda\simeq 0.22$ which, to a good
approximation, gives the Cabbibo mixing $|V_{us}|$.
Taking into account the experimental uncertainties that can be
as large as a factor of two one finds the following mass patterns
for the up and down quarks:
\begin{equation}
m_u : m_c : m_t \sim ({\lambda^8\over\chi^3} : {\lambda^4\over\chi^3} : 1)
\times m_t
\end{equation}
\begin{equation}
m_d : m_s : m_b \sim ({\lambda^4\over\chi} : {\lambda^2\over\chi} : 1)
\times m_b \,.
\end{equation}
The parameter $\chi \sim 0.7$ gives the radiative
corrections (from the heavy top quark) to the mass ratios
when they run from the unification scale $M_G = 10^{16}$ GeV down to the
electroweak scale according to the renormalisation group equations
of the minimal supersymmetric standard model (MSSM). This effect
is particularly felt in the up-quark sector. The regularity of the
spectra in terms of $\lambda$ becomes more transparent at the very high
scales, where one also notices that ratios of up quark masses and down
quark masses are related through a $\lambda^2 \leftrightarrow \lambda$
transformation. Compared to the almost equal spacing between neighbouring
quark mass levels the charged
lepton masses are perceived as an anomaly
\begin{equation}
m_e : m_{\mu} : m_{\tau} \sim (\lambda^5 :
\lambda : 1)\times m_{\tau}
\end{equation}
even at the unification
scale where apart from the unification of gauge couplings
one has also partial unification of Yukawa couplings
and the mass relations:
\begin{equation}
m_{\tau}\simeq m_b \qquad{\rm and}\qquad m_e \cdot m_{\mu}\simeq
m_d \cdot m_s  \quad .
\end{equation}

\noindent
Given the fact that at low energies one has only 13 observables (six quark and
three lepton masses, the three mixing angles and the CP violating phase
of the CKM matrix) one cannot fix the entries of
the quark and lepton mass matrices $M_u$, $M_d$ and $M_e$ at $M_G$,
even by assuming that the latter are symmetric.
This has led to different {\it Ans\"atze} [1,2] in which some
of the entries are zero while the others are given in powers of
$\lambda$. In the quark sector the maximum number of zeros that one can
have is five (counting together those in $M_u$ and $M_d$, but without
counting symmetric entries twice) and
there are five different pairs of $M_u - M_d$ textures at $M_G$
that lead to masses and mixings which are compatible with the
present-day experimental values [2].
Some extra consideration is thus needed to single out a
unique solution. In fact,
the zeros in the mass matrices
can be thought off as ``relics'' of a new symmetry which is not
``family-blind'', while the small non-zero entries could well be
correction terms generated after symmetry breaking.
I will discuss this ``old'' idea in the light of a new way of obtaining
also the successful $sin^2 \theta_W = 3/8$ result
of the canonical gauge coupling unification
which consists in extending the gauge group of the
standard model by a horizontal $U(1)_X$ factor whose
anomalies can be cancelled by a Green-Schwarz mechanism [3].

\noindent
Let us assume the existence of a family-dependent $U(1)_X$ gauge
symmetry at $M_{Planck}$, with respect to
which the quarks and leptons carry charges $\alpha_i$ and $a_i$
respectively, where $i=1,2,3$ is the generation index.
We first consider the up quark mass matrix.
Given the predominant role played by the top quark it is not unnatural
to start with a rank-one matrix and make a choice for the charges such
that only the (3,3) renormalizable coupling $t^c t h_1$ is allowed.
This fixes the charge of the light Higgs $h_1$ to $-2\alpha$
($\alpha\equiv \alpha_3$). We expect the other entries to be generated
by higher-dimension operators which may occur at
the string compactification level  and
contain combinations of scalar fields some of which acquire
vacuum expectation values (vev's) leading to spontaneous symmetry
breaking and large masses for the non-observable part of the spectrum.
For simplicity let us assume a pair of singlet fields $\sigma_{\pm}$
developping equal (vev's) along a ``D-flat'' direction and carrying
opposite charges $\pm 1$. They can give rise to higher-order couplings
$q^c_{i} h_1 ({<\sigma>\over M})^{|2\alpha - \alpha_i - \alpha_j|} q_{j}$.
Notice that when the exponent is positive (negative) only the field
$\sigma_+$ $(\sigma_-)$ can contribute. The new scale
${\cal E} = {<\sigma>\over M}$ which enters in the quark mass matrix is
the ratio of the symmetry breaking scale to the scale that governs these
higher-dimension operators, and could be the string unification scale
$M_S \simeq 10^{18}$ GeV or $M_{Planck}$. If ${\cal E}$ is a small
number one finds two universal hierarchy patterns in the generated texture:
\begin{equation}\label{Q0}
M_x \sim \left(
\begin{array}{ccc}
{\cal E}^{2 |x_1|} & {\cal E}^{|x_1 + x_2|} & {\cal E}^{|x_1|}\\
{\cal E}^{|x_1 + x_2|} & {\cal E}^{2 |x_2|} & {\cal E}^{|x_2|}\\
{\cal E}^{|x_1|} & {\cal E}^{|x_2|} & 1\\
\end{array}
\right) \qquad  |x_{1,2}| = |\alpha - \alpha_{1,2}| \,,
\end{equation}
namely $m_{11} \sim m_{13}^2$ and $m_{22} \sim m_{23}^2$.
The choice $|x_2|=1$ and $|x_1|=4$ or $|x_1|=2$
leads to:
\begin{equation}\label{xx}
M_u^F \sim \left(
\begin{array}{ccc}
{\cal E}^8 & {\cal E}^3 & {\cal E}^4\\
{\cal E}^3 & {\cal E}^2 & {\cal E}\\
{\cal E}^4 & {\cal E} & 1\\
\end{array}
\right) \quad {\rm or} \quad
M_u^G \sim \left(
\begin{array}{ccc}
{\cal E}^4 & {\cal E}^3 & {\cal E}^2\\
{\cal E}^3 & {\cal E}^2 & {\cal E}\\
{\cal E}^2 & {\cal E} & 1\\
\end{array}
\right) \,.
\end{equation}
If ${\cal E}$ is of order $\lambda^2$ the two textures above
correspond to the phenomenologically acceptable {\it Ans\"atze
\`a la Fritzsch} \footnote{The original {\it Ansatz} had a zero in
the (2,2) entry.}
or {\it \`a la Giudice} for the up quark mass matrix.
The generation of other acceptable textures having a zero in the (2,2)
or the (2,3) entry (but not in both entries simultaneously)
necessitates a more complicated mechanism involving
extra singlets and mixing with heavy Higgses.

\noindent
The next task is, given the up quark matrices of eq.(6), to construct
an acceptable down quark matrix. The assumption of symmetric mass
matrices and the $SU(2)_L$ symmetry require the equality
between the charges of the up and down quarks.
Assuming again that only the (3,3) renormalizable coupling is allowed
leads to the other light Higgs $h_2$ carrying the same charge as $h_1$.
This means that this $U(1)_X$ is anomalous and needs a cancellation
mechanism which will be discussed at the end of the talk.
Notice that the choice of a particular texture for $M_u$ has already
fixed the texture of $M_d$ in terms of some new scale
${\cal E}^{\prime}$ which has to be of order $\lambda$ to give the
correct mass spectrum of eq.(2). The origin of this difference in scale
${\cal E}^{\prime}\sim {\cal E}^{1/2}$ is yet unknown.

\noindent
Thus if one insists in generating the up and down quark mass matrices
through the same (simple) mechanism that led to
eq.(5) one can only generate textures {\it \`a la Fritzsch} containing
four zeros in total. In order to generate the five-zero textures
of ref.[2] some sort of extension is needed.
In many compactification schemes there are among other things
additional pairs of  Higgs fields which acquire masses at the scale of
symmetry breaking.
One can thus envisage the possibility of mixing between the light Higgses
$h_{1,2}$ with heavy Higgses $H_{1,2}$ whose charge we denote
by $-2\beta$. Then one can also have couplings
$q^c_{i} H ({<\sigma>\over M})^{|2\beta - \alpha_i - \alpha_j|} q_{j}$
which give rise to the following texture:
\begin{equation}\label{Yz}
M_z \sim \left(
\begin{array}{ccc}
{\cal E}^{2 |z_1|} & {\cal E}^{|z_1 + z_2|}
& {\cal E}^{|z_1 + z|}\\
{\cal E}^{|z_1 + z_2|} & {\cal E}^{2 |z_2|}
& {\cal E}^{|z_2 + z|}\\
{\cal E}^{|z_1 + z|} & {\cal E}^{|z_2 + z|}
& 1 + {\cal E}^{2 |z|}\\
\end{array}
\right) \,,
\end{equation}
whith $|z_i| = |\beta - \alpha_i|$, and
$|z| = |\beta - \alpha|$.
When the difference between the light- and heavy-Higgs charges is larger
than between the charges of the heavy Higgs and the quarks this
automatically gives suppressed (1,3) and (2,3) mass entries.
Taken together, the textures of eqs.(5,7) can generate any acceptable up
or down quark texture. On the other hand the generation of an
acceptable set of up and down quark matrices with five zeros is again too
constrained, while the {\it Fritzsch Ansatz} can be
obtained also from the conditions $|x_2| = 1$, $|z_1+z_2| = 3$ and
$|x_1|, |z_{1,2}|, |z|\gg 0$ [4].

\noindent
What about the lepton sector?
Assuming simply the gauge symmetries of the SM the $U(1)_X$ charges
$a_i$ of the
leptons  are not related to those of the quarks.
Allowing however the coupling $\tau^c \tau h_2$ leads to $a_3 = \alpha$
and to $m_{\tau} = m_b$ unification.
Another constraint comes from the second mass relation
of eq.(4) which means that we should be looking for textures for which
det $M_e$ = det $M_d$ and therefore for combination of charges that
satisfy: $a_1 + a_2 = \alpha_1 + \alpha_2$.
Since the early days of grand unification it is known
that in order to obtain also for the first two generations
acceptable mass relations between the charged leptons and down quarks
the (2,2) entry of $M_d$ should be multiplied by a factor of minus
three, the other entries of $M_e$ and $M_d$ been equal.
In this way, though there can be no explanation of the factor minus three
in this approach, in terms of the scale ${\cal E}^{\prime}$ and its
various powers, $M_e$ becomes identical with $M_d$.
The alternative is a texture:
\begin{equation}\label{Q0}
M_e \sim \left(
\begin{array}{ccc}
0 & \lambda^3 & 0\\
\lambda^3 & \lambda & 0\\
0 & 0 & 1\\
\end{array}
\right)  \,,
\end{equation}
which can be generated from the texture $M_z$ in eq.(7) from the choices
$|z_2| = 1/2$ and $|z_1| = 5/2$ or $7/2$ when $|z|\gg |z_{1,2}|$.

\noindent
We turn now to the neutrino sector.
Again as a consequence of the $SU(2)_L$ symmetry and our symmetric
{\it Ansatz} the lefthanded and righthanded neutrinos, $\nu_i$ and
$N_i$, become charged under the $U(1)_X$ with the same charges $a_i$
as the charged leptons. Obviously the presence
of the $N_i$'s implies a larger symmetry than what has been assumed so far,
but also the assumption of symmetric mass matrices can find its
justification only in the context of a left-right symmetric
theory. Starting from $M_e = M_d$ and
assuming that the same mechanism which generated masses for the
quarks and the leptons generates also Dirac masses for the neutrinos
we are automatically led to another well known GUT equality:
$M_{\nu}^D = M_u$. In the same way starting from eq.(8) one obtains
$M_{\nu}^D = M_e$.
On the other hand, Majorana mass terms $M_R^{ij}$ need not be
generated in the same way. In compactified string models, due
to the absence of large Higgs representations, righthanded
neutrinos donot get tree-level masses, so all entries
in $M_R$ are due to nonrenormalizable operators,
and nothing is a priori known concerning the particular
texture of $M_R$ or the existence of a possible
hierarchy
in this sector. The only constraints
come from the requirement that the seesaw-suppressed masses of
the ordinary neutrinos should be below the experimental upper limits.
For this, $M_R$ has to be a nonsingular matrix and its scale
should be well above the electroweak scale.
Therefore in addition to the operators that generated the textures
of eqs.(5,7) one will need at least an extra piece to set the Majorana
mass scale. Common examples are operators containing the heavy Higgses
which have been used for generating the texture in eq.(7),
namely $N_i^c H H N_j$,
whose scale is of ${\cal O}(M_G^2/M_S)$ multiplied for some orbifold
suppression factor. In this case the condition
$\alpha_i + \alpha_j = 4\beta$ leads to
$M_R^{ij}\not= 0$. One can always fix the charges of the light and heavy
Higgses relative to each other such that there is at least one entry which is
different from zero.
On the other hand, the charges $\alpha_1,\alpha_2$ (and $a_1,a_2$) are
fixed
by the conditions which led to eqs.(6,8) in terms of the Higgs charges.
Therefore there is no freedom in generating more entries through the
same operator. There are two alternatives paths leading to the
generation of a nonsingular $M_R$:
Either some of the
Majorana entries are generated perturbatively in a similar way as the
entries in the other mass matrices (this implies a hierarchy
of righthanded neutrino scales) or one can assume a different set of
singlets in the higher-dimension operators which give rise to the
texture $M_z$ in eq.(7).
This freedom allows the generation of $M_R$
textures containing four zeros [4] and leading to interesting
predictions for neutrino oscillation experiments [5].

\noindent
Let us finally comment on the cancellation of the mixed
anomalies of a $U^{\prime}(1)$ with
$SU(3)$, $SU(2)$ and $U(1)_Y$
via a Green-Schwarz (GS) term in 4D string theories.
It was pointed out that the existence of an axion with
universal couplings to all gauge groups together
with the assumption that the gauge couplings, up
to normalisation factors $k_i$, meet at some scale,
can lead to cancellation of the anomalies
by an appropriate shift of the axion field
but also to gauge coupling unification
if the anomaly coefficients are in the ratio
$A_3 : A_2 : A_1 = k_3 : k_2 : k_1 = 1 : 1 : 5/3$ [6].
It is remarkable that this seems to be the case also
for the $U(1)_X$ that gave an acceptable mass spectrum [3].

\end{document}